\documentclass[11pt]{article}
\usepackage{amssymb,amsmath,amsthm,amscd,bbm}
\usepackage{amsfonts}

\def\a{\alpha}
\def\b{\beta}
\def\g{\gamma}

\def\ve{\varepsilon}

\def\h{\eta}
\def\t{\theta}
\def\vt{\vartheta}

\def\vp{\varphi}
\def\c{\raise2pt\hbox{$\chi$}}

\def\sfrac#1#2{{\textstyle\frac{#1}{#2}}}

\def\+{\dagger}
\def\={\ =\ }
\def\pa{\partial}

\def\>{\rangle}
\def\<{\langle}

\newcommand{\e}{\,\mathrm{e}\,}
\newcommand{\im}{\,\mathrm{i}\,}
\newcommand{\diff}{\mathrm{d}}
\newcommand{\R}{{\mathbb{R}}}
\newcommand{\C}{{\mathbb{C}}}

\newcommand{\unity}{\mathbbm{1}}
\newcommand{\cH}{{\mathcal H}}

\newcommand{\beq}{\begin{equation}}
\newcommand{\eeq}{\end{equation}}
\newcommand{\bea}{\begin{eqnarray}}
\newcommand{\eea}{\end{eqnarray}}
\newcommand{\bal}{\begin{aligned}}
\newcommand{\eal}{\end{aligned}}

\newcommand{\und}{\qquad\text{and}\qquad}

\begin{document}
\begin{titlepage}
\setcounter{page}{0}
\begin{flushright}
hep-th/0507062\\
ITP--UH--10/05
\end{flushright}

\vspace{10mm}

\begin{center}

{\Large\bf  Time-Space Noncommutative Abelian Solitons}

\vspace{20mm}

{\Large Chong-Sun Chu~$^1$\ \ and \ \ Olaf Lechtenfeld~$^2$} \\[24pt]

{\em ${}^1\ $ 
     Centre for Particle Theory and Department of Mathematics \phantom{X}\\
     University of Durham, Durham, DH1 3LE, United Kingdom}\\
{\tt chong-sun.chu@durham.ac.uk} \\[12pt]     

{\em ${}^2\ $
     Institut f\"ur Theoretische Physik, Universit\"at Hannover \phantom{X}\\
     Appelstra\ss{}e 2, D-30167 Hannover, Germany}\\
{\tt lechtenf@itp.uni-hannover.de}

\vspace{30mm}

\begin{abstract}
\noindent 

We demonstrate the construction of solitons for a time-space Moyal-deformed
integrable U($n$) sigma model (the Ward model) in $2{+}1$ dimensions.
These solitons cannot travel parallel to the noncommutative spatial direction.
For the U(1) case, the rank-one single-soliton configuration is constructed
explicitly and is singular in the commutative limit. The projection to $1{+}1$
dimensions reduces it to a noncommutative instanton-like configuration.
The latter is governed by a new integrable equation, which describes
a Moyal-deformed sigma model with a particular Euclidean metric and a
magnetic field.

\end{abstract}

\end{center}

\vfill

\end{titlepage}

\section{Introduction and results}

\noindent
According to Ward's conjecture~\cite{ward1} all integrable equations can be
obtained by dimensional and algebraic reduction of the self-dual Yang-Mills
(SDYM) equations in four dimensions. This supposition extends to the Moyal-Weyl
deformation of integrable systems~\cite{hamtoda}. 
Indeed, maximal integrability seems to be guaranteed only for those
noncommutative equations which descend from SDYM theories on
noncommutative $\R^{4,0}$ or $\R^{2,2}$~\cite{ncsg}.

When the mother theory is non-abelian, the solitonic solutions of the reduced
model connect smoothly  to their undeformed cousins in the commutative limit.
If one starts with the deformed U(1) SDYM equations, however, this will no 
longer be the case, because the latter trivializes in the commutative limit. 
Instead, noncommutative solitons stemming from an abelian mother theory
must have a singular commutative limit, analogous to the celebrated 
noncommutative abelian instantons~\cite{ncinst}.

A known example of this phenomenon occurs for the $2{+}1$ dimensional 
Yang-Mills-Higgs system whose solitons are provided by a modified sigma model 
(the Ward model~\cite{ward2}). For the space-space noncommutative 
Ward model, multi-solitons have been constructed and studied~\cite{ncward}
by passing to the operator formalism using the Moyal-Weyl map.
The subclass of static abelian solitons solves the noncommutative standard
sigma-model equation in $2{+}0$ dimensions~\cite{domrin}. For the time-space 
noncommutative case, much less is known, partially due to the lack of an 
apparent operator formalism. For instance, the reduction of the U(2)~Ward model
to various noncommutative sine-Gordon systems in $1{+}1$ dimensions has been 
investigated~\cite{ncsg}. For other examples and further references, 
see e.g.~\cite{hamanaka}. However, the U(1) case remains illusive so far.

Therefore, the question arises as to whether there exist {\it time-space\/} 
noncommutative integrable equations which stem from the {\it abelian\/} SDYM
theory on the Moyal-deformed $\R^{2,2}$. 
To proceed, one may step down to the $2{+}1$ dimensional
time-space noncommutative U(1) Ward model and look there for specific 
classical solutions which have a chance to descend to noncommutative 
abelian solitons in $1{+}1$ dimensions. Equivalently, one may search for 
an integrable dimensional reduction of the deformed abelian Ward model's 
equation of motion, in order to obtain an integrable system in two dimensions
featuring time-space noncommutative solitons with a singular commutative limit.
In this letter, we will follow this strategy to construct the desired 
$1{+}1$ dimensional integrable system. 

We summarize the organization and results of the letter as follows.
In the next section, we give a brief review of the Ward model. 
Next, we demonstrate how to construct
multi-center solutions of the time-space noncommutative U($n$) Ward model 
by employing a linear system and the so-called dressing method~\cite{dressing}.
Important to our construction is the observation that, despite the fact that
time is noncommuting, a specific co-moving coordinate system can be
introduced such that the coordinate commutators form the standard Heisenberg 
algebra, and hence a Fock-space construction is possible.
The rank-one single-soliton configuration is written down explicitly, 
and its properties are discussed. In particular, we find that the construction 
degenerates when the soliton moves in the deformed spatial direction. 
We remark that the Ward soliton satisfies a BPS-type equation and enjoys 
the standard shape invariance which implies a time-like translational 
symmetry. Therefore, it depends only on two of the three co-moving coordinates.
This fact allows us to perform a dimensional reduction in the symmetry 
direction to obtain a $1{+}1$ dimensional noncommutative system.
On the Ward soliton, this reduction takes a twisted form and produces an
instanton-like configuration in the Moyal-deformed two-dimensional Minkowski
space. The corresponding novel integrable equation of motion employs 
a Euclidean metric although we start out in Minkowski space. This flip of 
signature is a generic feature of the integrable reduction in the one-soliton
sector.

\section{The Ward model}

\noindent
The U($n$) Yang-Mills-Higgs equations in $2{+}1$ dimensions are implied by 
the Bogomolnyi equations
\beq \label{bog}
\sfrac12\ve^{abc} (\pa_{[b}A_{c]}+A_{[b}A_{c]}) \=
\pa^a\vp + [A^a,\vp] \qquad\text{with}\quad a,b,c\in\{0,1,2\}
\eeq
for the gauge potential~$A=A_a\diff x^a$ and the Higgs field~$\vp$, 
all Lie-algebra valued. Contractions employ the Minkowski metric
$(\h_{ab})=\text{diag}(-1,+1,+1)$.
With the ansatz and light-cone gauge-fixing
\beq
\bal
A_t - A_y \= 0  \quad&,\qquad
A_t + A_y \= \Phi^\+ (\pa_t + \pa_y) \Phi \quad, \\
A_x + \,\vp\, \= 0  \quad&,\qquad
A_x - \,\vp\, \= \Phi^\+ \pa_x \Phi \quad,
\eal
\eeq
introducing a U($n$)-valued prepotential~$\Phi$,
the Bogomolnyi equations~(\ref{bog}) reduce to the Yang-type equation
\beq \label{wardeq1}
-\pa_t(\Phi^\+\pa_t\Phi) + \pa_x(\Phi^\+\pa_x\Phi) + \pa_y(\Phi^\+\pa_y\Phi)
+\pa_y(\Phi^\+\pa_t\Phi) - \pa_t(\Phi^\+\pa_y\Phi) \= 0 \quad,
\eeq
making explicit the breaking of $SO(2,1)$ Lorentz invariance by the ansatz.
This can be more succinctly written as
\beq \label{wardeq2}
(\h^{ab} + k_c \ve^{cab}) \pa_a (\Phi^\+\pa_b\Phi) \= 0
\qquad\text{for}\quad (k_c)=(0,1,0) \quad,
\eeq
which is the equation of motion for a (WZW-modified)
nonlinear sigma model known as Ward's model~\cite{ward2}. 
The modification may be interpreted as a background magnetic field
$b^{ab}=k_c\ve^{cab}$ in the $k$~direction.
This model is integrable at the price of foregoing Lorentz invariance.

It is convenient to introduce light-cone coordinates
\beq
u\ :=\ \sfrac{1}{2}(t+y)\quad,\qquad
v\ :=\ \sfrac{1}{2}(t-y)\quad,\qquad
\pa_u\ =\ \pa_t+\pa_y\quad,\qquad
\pa_v\ =\ \pa_t-\pa_y\quad,
\eeq
in terms of which the Ward equation~(\ref{wardeq1}) simplifies to
\beq \label{wardeq3}
\pa_x(\Phi^\+\pa_x\Phi) - \pa_v(\Phi^\+\pa_u\Phi) \= 0 \quad.
\eeq

\section{Linear system and single-pole ansatz}
\noindent
It is easy to see that~(\ref{wardeq3}) is the compatibility condition of
two linear equations,
\beq
\label{linsys}
(\zeta \pa_x -\pa_u)\psi\ =\ (\Phi^\+\pa_u\Phi)\,\psi 
\qquad\textrm{and}\qquad
(\zeta \pa_v -\pa_x)\psi\ =\ (\Phi^\+\pa_x\Phi)\,\psi \quad,
\eeq
which can be obtained from the Lax pair for the SDYM equations in $\R^{2,2}$ 
\cite{IP2} by gauge-fixing and dimensional reduction.
Here, $\psi$ is U($n$) valued and depends on $(x,u,v,\zeta)$.
The spectral parameter~$\zeta$ lies in the extended complex plane~$\C P^1$.
The auxiliary function $\psi$ is subject to the reality condition~\cite{ward2}
\beq \label{real}
\psi(x,u,v,\zeta)\;[\psi(x,u,v,\bar{\zeta})]^{\dagger}\ =\ \unity \quad.
\eeq
We also impose on $\psi$ the standard asymptotic conditions 
\beq \label{asymp}
\psi(x,u,v,\zeta\to\infty)\ =\ \unity \und
\psi(x,u,v,\zeta\to0)\ =\ \Phi^\+(x,u,v) \quad.
\eeq
Once we have a solution $\psi$ to~(\ref{linsys}) the prepotential $\Phi$ 
may be reconstructed from its asymptotic value via~(\ref{asymp}), but the
gauge potentials $A_u=\Phi^\+\pa_u\Phi$ and $A_x{-}\vp=\Phi^\+\pa_x\Phi$ 
can also be found from
\bea \label{diff1}
-\psi(x,u,v,\zeta)\;(\zeta\pa_x-\pa_u)[\psi(x,u,v,\bar{\zeta})]^{\dagger}\
=\ (\Phi^\+\pa_u\Phi)(x,u,v) \quad, \\[4pt] \label{diff2}
-\psi(x,u,v,\zeta)\;(\zeta\pa_v-\pa_x)[\psi(x,u,v,\bar{\zeta})]^{\dagger}\
=\ (\Phi^\+\pa_x\Phi)(x,u,v) \quad.\,
\eea

A nontrivial solution~$\psi$ holomorphic in~$\zeta$ cannot be analytic
on~$\C P^1$, hence must have poles, say at $\zeta=\mu_k\in\C$ 
with $k=1,\dots,m$.
Contrasting the ensueing $\zeta$~dependence of the left-hand sides of
(\ref{real}), (\ref{diff1}) and (\ref{diff2}) with the $\zeta$~independence
of their right-hand sides, consistency requires that the residues at
$\zeta{=}\mu_k$ and $\zeta{=}\bar\mu_k$ must vanish. In fact this condition
also suffices to guarantee the existence of a solution.
It is known that such solutions describe multi-soliton configurations, 
with the velocities of the individual solitons for $\mu{=}\mu_k$ being given by
\beq \label{vel}
(v_x,v_y) \= -\Bigl(
\frac{\mu+\bar\mu}{\mu\bar\mu+1}\ ,\ \frac{\mu\bar\mu-1}{\mu\bar\mu+1}\Bigr)\=
-\bigl(\mu-\bar\mu-\sfrac{1}{\mu}+\sfrac{1}{\bar\mu}\bigr)^{-1}\, \Bigl( 
\frac{\mu}{\bar\mu}-\frac{\bar\mu}{\mu}\ ,\ 
\mu-\bar\mu+\frac{1}{\mu}-\frac{1}{\bar\mu}\Bigr) \quad.
\eeq
This relation carries over to the noncommutative realm~\cite{ncward}.

In this letter we only consider the simplest case of a single soliton,
i.e.~$m{=}1$, and comment on its generalization at the end. Accordingly, 
we make the single-pole ansatz (dropping the index~$k$)
\beq \label{ansatz}
\psi(x,u,v,\zeta) \= \unity + \frac{\mu{-}\bar\mu}{\zeta{-}\mu}P(x,u,v)\quad.
\eeq
The group-valued function~$P$ is determined from the zero-residue conditions
\bea
(\unity-P)\,P^\+ &=& 0 \quad, \label{proj} \\
(\unity-P)\,(\bar\mu\pa_x - \pa_u)\,P^\+ &=& 0 \quad,\label{res1}\\
(\unity-P)\,(\bar\mu\pa_v - \pa_x)\,P^\+ &=& 0  \label{res2}
\eea
and their hermitian conjugates.
It follows from~(\ref{proj}) that~$P$ is a hermitian projector,
\beq \label{hermproj}
P^2 \= P \= P^\+ \quad,
\eeq
so that the two differential operators in (\ref{res1}) and (\ref{res2})
are seen to map the image of~$P$ into itself.

\section{Co-moving coordinates}
\noindent
The remaining conditions (\ref{res1}) and (\ref{res2}) restrict the
coordinate dependence of~$P$. They can be simplified with an appropriate
linear coordinate transformation $(u,v,x)\mapsto(w,\bar w,s)$. In
general, the coordinate transformation takes the form
\beq
w\ :=\ \nu\,x + z \quad,\qquad 
\bar w\ :=\ \nu\,x + \bar z\quad,\qquad 
s\ :=\ c\,x + \g\,z + \bar \g\,\bar z \quad,
\eeq
where the complex combinations 
\bea
z\ :=\ \a\,u + \b\,v \und \bar z\ :=\ \bar \a\,u + \bar \b\,v
\eea
are introduced for later convenience (see section 6),
and the coefficients $\nu, \a,\b, \g \in \C$ and $c \in \R$
are to be chosen. In terms of the new coordinates, 
the differential operators in (\ref{res1}) and~(\ref{res2}) become
\beq \label{del}
\bal
\pa_x - \sfrac{1}{\bar \mu}\,\pa_u &\= 
(\nu - \sfrac{\a}{\bar \mu})\,\pa_w + 
(\bar \nu - \sfrac{\bar\a}{\bar \mu})\,\pa_{\bar w} +
(c- \sfrac{\g\a}{\bar \mu} - \sfrac{\bar \g \bar \a}{\bar \mu})\,\pa_s \quad,
\\[4pt]
\pa_x - \bar \mu\,\pa_v &\=
(\nu - \b \bar \mu)\,\pa_w + 
(\bar \nu - \bar \b \bar \mu)\,\pa_{\bar w} +
(c- \g \b \bar \mu - \bar \g \bar \b \bar \mu)\,\pa_s \quad. 
\eal
\eeq
In general, $P$ depends on $w$, $\bar w$ and $s$.  
As we will see, the system (\ref{res1}), (\ref{res2}) becomes exactly solvable
if there is a translational symmetry. In our case with noncommutativity,
and with the goal to look for a Fock-space construction of solutions,  
we take this symmetry to be generated by $\pa_s$
so that our solutions depend on the pair $(w,\bar w)$ alone.
In this case, nontrivial solutions can be obtained if the two 
differential operators in (\ref{del}) with $\pa_s{=}0$ become collinear.
Without loss of generality, 
this happens when the coefficients of~$\pa_w$ vanish, i.e.~for
\beq
\a = \nu \bar \mu \und \b = \sfrac{\nu}{\bar \mu} 
\qquad\text{with \quad $\nu$ left free} \quad.
\eeq

For convenience, we choose the time-like vector $\pa_s$
to be normalized to one and orthogonal to the $w\bar w$ plane,
which fixes the coefficients $c$ and~$\g$.
Our coordinate transformation then reads
\beq \label{wdef}
\bal
w\ &=\quad \nu\ \bigl[\bar{\mu}\,u +\sfrac{1}{\bar\mu}\,v + x \bigr] 
\qquad\quad\ \ = \quad\;
\nu\ \bigl[x + \sfrac{1}{2}(\bar{\mu}{-}\sfrac{1}{\bar\mu})\,y +
\sfrac{1}{2}(\bar{\mu}{+}\sfrac{1}{\bar\mu})\,t \bigr] \quad, \\[4pt] 
\bar w\ &=\quad \bar\nu\ \bigl[\mu\,u +\sfrac{1}{\mu}\,v + x \bigr] 
\qquad\quad\ \ = \quad\;
\bar\nu\ \bigl[x + \sfrac{1}{2}(\mu{-}\sfrac{1}{\mu})\,y +
\sfrac{1}{2}(\mu{+}\sfrac{1}{\mu})\,t \bigr] \quad, \\[4pt]
s \ &=\;\sfrac{-2\im}{\mu-\bar\mu}
\bigl[\mu\bar\mu\,u + v + \sfrac12(\mu{+}\bar\mu)\,x \bigr] \ =\; 
\sfrac{-\im}{\mu-\bar\mu}
\bigl[(\mu{+}\bar\mu)\,x +(\mu\bar\mu{-}1)\,y +(\mu\bar\mu{+}1)\,t
\bigr]\quad.
\eal
\eeq
For the vector fields we arrive at
\beq \label{pawdef}
\bal
\pa_w &\= \sfrac{1}{\nu} \sfrac{\mu\bar\mu}{(\mu-\bar\mu)^2}\,
\bigl[\sfrac{1}{\mu}\,\pa_u + \mu\,\pa_v -2\,\pa_x \bigr] 
\quad,\qquad\qquad
\pa_u \= \nu\bar\mu\,\pa_w + \bar\nu\mu\,\pa_{\bar w} 
- \sfrac{2\im \mu\bar\mu}{\mu-\bar\mu} \,\pa_s 
\quad, \\[4pt]
\pa_{\bar w} &\= \sfrac{1}{\bar\nu} \sfrac{\mu\bar\mu}{(\mu-\bar\mu)^2}\,
\bigl[\sfrac{1}{\bar\mu}\,\pa_u + \bar\mu\,\pa_v -2\,\pa_x \bigr] 
\quad,\qquad\qquad
\pa_v \= \sfrac{\nu}{\bar\mu}\,\pa_w+\sfrac{\bar\nu}{\mu}\,\pa_{\bar w}
- \sfrac{2\im}{\mu-\bar\mu}\,\pa_s
\quad, \\[4pt]
\pa_s &\ = \qquad\sfrac{-\im}{\mu-\bar\mu}\,
\bigl[\pa_u + \mu\bar\mu\,\pa_v -(\mu{+}\bar\mu)\,\pa_x\bigr]
\quad,\qquad
\pa_x \= \nu\,\pa_w + \bar\nu\,\pa_{\bar w} 
- \sfrac{\im (\mu{+}\bar\mu)}{\mu -\bar\mu}\,\pa_s \quad,
\eal
\eeq
revealing degeneracy for $\mu\in\R$. The Euclidean metric induced 
on the $w\bar w$ plane is proportional to $(\text{Im}\,\mu)^{-2}$.
We remark that, up to normalization, the expression for $\pa_s$ 
is independent of the value taken for $c$ or $\g$ in~(\ref{del}), 
so that the choice of~$s$ does not matter for the rest of the letter.

Looking at~(\ref{vel}), we notice that 
\beq
\text{Im}\,\mu=0 \quad\Leftrightarrow\quad v_x^2+v_y^2=1 \und
\mu=\pm\im \quad\Leftrightarrow\quad v_x=v_y=0 \quad,
\eeq
so the transformation~(\ref{wdef}) degenerates for light-like solitons.
Let us take Im$\,\mu<0$ from now on.
Note that for static solitons ($\mu=-\im$) the above relations simplify to
\beq
w \= \nu\,(x + \im\,y) \quad,\qquad
\bar w \= \bar\nu\,(x - \im\,y) \quad,\qquad 
s \= t \quad.
\eeq

In terms of the new coordinates, the differential equations 
(\ref{res1}) and~(\ref{res2}), with (\ref{hermproj}), combine to
\beq \label{resnew}
(\unity-P)\,\pa_{\bar w}\,P \= 0 \und
(\unity-P)\,\pa_s\,P \= 0 \quad.
\eeq
As indicated above, we are interested in $s$~independent solutions. 
Hence, one is looking 
for projectors $P(w,\bar w)$ satisfying a single ``BPS'' equation
\beq \label{BPS}
\pa_{\bar w} P \= P\,\gamma \qquad\text{for some}\quad \gamma \quad.
\eeq
Physically, 
although the solitons of the Ward model are neither Galilei nor Lorentz 
invariant, they depend -- like all solitons -- only on their moving-frame
coordinates. The combinations $(w,\bar w)$ are precisely those co-moving 
coordinates. Consequently, the vector field $\pa_s$ generates a symmetry 
of the one-soliton configuration, i.e.~$\pa_s P=0$. 
An obvious solution to~(\ref{BPS}) is to build~$P$ from a holomorphic
function~$T(w)$ which spans the image of~$P$.
This kind of solution extends to the noncommutative case 
when the gauge group is nonabelian. In the abelian situation, however, 
only genuinely noncommutative solutions exist, which we will discuss now.

\section{Time-space noncommutativity and abelian solitons}
\noindent

Up to this point we have basically followed \cite{ward2} which discusses
the commutative case. The space-space noncommutative deformation was analyzed
in~\cite{ncward}, with
\beq
[x,y] \= \im\theta 
\qquad\Longrightarrow\qquad [w,\bar w] \= 2\,\theta\ >0
\qquad\text{for}\quad
\nu\bar\nu=-4\im(\mu{-}\bar\mu{-}\sfrac{1}{\mu}{+}\sfrac{1}{\bar\mu})^{-1} 
\quad.
\eeq
Due to the anisotropy of the Ward model, there are many inequivalent choices
of time-space deformations. The generic cases are~\footnote{
Note that $[w,s]\neq 0$. For example, for the case (\ref{ty-alg}), 
one has $[w,s]=\nu\,\theta$. However, since we are looking for 
$s$~independent solutions, we can safely ignore this noncommutative structure.
}
\bea
[t,x] \= \im\theta 
\qquad&\Longrightarrow&\qquad [w,\bar w] \= 2\,\theta\ >0
\qquad\text{for}\quad
\nu\bar\nu=\ \ 4\im(\mu{-}\bar\mu{+}\sfrac{1}{\mu}{-}\sfrac{1}{\bar\mu})^{-1}
\quad. \label{tx-alg} \\[4pt]
[t,y] \= \im\theta
\qquad&\Longrightarrow&\qquad [w,\bar w] \= 2\,\theta\ >0
\qquad\text{for}\quad
\nu\bar\nu=-4\im(\sfrac{\mu}{\bar\mu}-\sfrac{\bar\mu}{\mu})^{-1}
\quad. \label{ty-alg}
\eea
Interestingly, the first case becomes singular for $|\mu|=1$, i.e.~$v_y=0$,
while for the second case this happens when Re$\,\mu=0$, i.e.~$v_x=0$.
More generally, we find that
\beq
[t\,,\,x\cos\vt+y\sin\vt] \= \im\theta \qquad\Longrightarrow\qquad 
[w,\bar w]\ \propto\ \nu\bar\nu\,(v_y\cos\vt - v_x\sin\vt) \quad.
\eeq
Apparently, solitonic motion in the direction of the deformed spatial
coordinate is prohibited, ruling out static solutions in particular.  

In order to describe abelian solitons, we specialize the target to~U($n{=}1$).
It is convenient to employ the operator formalism, which realizes the
noncommutativity via operator-valuedness in an auxiliary Fock space~$\cH$.
Then, the image of the projector~$P$ in the single-pole ansatz~(\ref{ansatz}) 
is spanned by a collection (row vector) of kets,
\beq \label{TandP}
|T\> \= \bigl( \ |T^1\>\quad |T^2\> \quad\ldots\quad |T^r\> \ \bigr)
\qquad\text{so that}\qquad P \= |T\>\<T|
\eeq
with $r=\text{rank}(P)$ and normalization $\<T|T\>=\unity_{r\times r}$.
Next, the BPS equation~(\ref{BPS}) reduces to
\beq
\pa_{\bar w} |T\> \= |T\>\,\Gamma 
\qquad\text{for some $r{\times}r$ matrix $\Gamma$} \quad.
\eeq
If we take
\beq
\Gamma \= \text{diag}\bigl( \a^1,\a^2,\ldots,\a^r \bigr) 
\qquad\text{with}\quad \a^i\in\C \quad,
\eeq
then we arrive at the conditions
\beq 
\pa_{\bar w} |T^i\> \= \a^i\,|T^i\> 
\qquad\text{for}\quad i=1,\ldots,r \quad.
\eeq
This identifies the kets as coherent states
\beq \label{coh1}
|T^i\> \= \e^{\a^i\bar w -\bar \a^i w}\,|0\> \ =:\ |\a^i\> 
\eeq
based on the Fock vacuum of the algebra (\ref{tx-alg}) or~(\ref{ty-alg}): 
\beq \label{wvac}
w\,|0\> \= 0 \quad.
\eeq
The rank-one solution reads (dropping the superscript)
\beq \label{Phisol}
\Phi \= \unity_{\cH} - (1{-}\sfrac{\mu}{\bar\mu})P
\= \unity_{\cH} - (1{-}\sfrac{\mu}{\bar\mu})|\a\>\<\a| 
\= \e^{\a\bar w-\bar\a w}
   \Bigl[ \unity_{\cH} - (1{-}\sfrac{\mu}{\bar\mu})|0\>\<0| \Bigr]
   \e^{\bar\a w-\a\bar w} \quad.
\eeq
At this point, it is instructive to recall that the initial problem with 
time-space noncommutativity was the lack of a Fock-space formalism. 
We circumvented the problem by passing to an appropriately defined 
co-moving coordinate system. Although the commutation relations 
take a more complicated form as a whole, they become simpler in a 
two~dimensional subspace and in fact take on the standard Heisenberg algebra 
form enabling a Fock-space construction.

To reveal the coordinate dependence of our soliton (\ref{Phisol})
and interpret the moduli~$\a$, we pass to the star-product formulation
using the Moyal-Weyl map,\footnote{
Note that not the star-exponential but the ordinary exponential appears below.}
\beq \label{starsol}
\Phi_\star \= 1 - 2(1{-}\sfrac{\mu}{\bar\mu}) \e^{-|w-2\t\a|^2/\t}
\= 1 - 2(1{-}\sfrac{\mu}{\bar\mu})\exp\Bigl\{ -\frac{1}{\t}\,
\bigl|\nu x + \sfrac{\nu}{2}(\bar{\mu}{-}\sfrac{1}{\bar{\mu}})\,y +
\sfrac{\nu}{2}(\bar{\mu}{+}\sfrac{1}{\bar{\mu}})\,t - 2\t\a\bigr|^2 \Bigr\}
\quad.\eeq
This is a Gaussian profile which peaks at
\beq
\bal 
x(t) &\= \bigl( \mu{-}\bar\mu{-}\sfrac{1}{\mu}{+}\sfrac{1}{\bar\mu} \bigr)^{-1}
\Bigl[ 2\t \bigl( \sfrac{\mu\a}{\nu}{-}\sfrac{\bar\mu\bar\a}{\bar\nu}
{-}\sfrac{\a}{\mu\nu}{+}\sfrac{\bar\a}{\bar\mu\bar\nu} \bigr)
- \bigl( \sfrac{\mu}{\bar\mu}{-}\sfrac{\bar\mu}{\mu} \bigr)\,t \Bigr] \quad, \\
y(t) &\= \bigl( \mu{-}\bar\mu{-}\sfrac{1}{\mu}{+}\sfrac{1}{\bar\mu} \bigr)^{-1}
\Bigl[ 4\t \bigl( \sfrac{\bar\a}{\bar\nu}{-}\sfrac{\a}{\nu} \bigr)
-\bigl(\mu{-}\bar\mu{+}\sfrac{1}{\mu}{-}\sfrac{1}{\bar\mu}\bigr)\,t\Bigr]\quad.
\eal
\eeq
Even without computing its energy density,
it is clear that this lump moves with constant velocity across the $xy$ plane. 
The velocity is determined by~$\mu$ as given in~(\ref{vel}), 
the zero-time position by~$\a$.
When the velocity points in the deformed direction, the Euclidean 
$w\bar w$~plane collapses and we lose the operator formalism.

\section{Reduction to two dimensions}
\noindent
Although the Ward model lives in $2{+}1$ dimensions, our noncommutative abelian
soliton~(\ref{Phisol}) is effectively a two-dimensional configuration, since
it does not depend on the $s$~coordinate. For a commutative soliton this is 
by no means surprising but a direct consequence of its shape invariance.
In the co-moving frame, where the soliton is at rest, it is obvious that this
symmetry is just time translation.
Despite the absence of rest frames in the time-space Moyal-deformed
Ward model, it remains true that the one-soliton sector enjoys the 
translational invariance generated by~$\pa_s$. 
This yields a reduced coordinate dependence for the one-soliton configuration:
Putting $\pa_s=0$ in~(\ref{pawdef}), we learn that
\beq \label{parel}
\pa_u + \mu\bar\mu\,\pa_v -(\mu{+}\bar\mu)\,\pa_x \= 0
\qquad\Longleftrightarrow\qquad
\pa_x \=
\nu\,\pa_w + \bar\nu\,\pa_{\bar w} \quad,
\eeq
which is actually obeyed on our solution~(\ref{starsol}).

What happens if we perform a reduction to $1{+}1$ dimensions,
retaining the time-space Moyal deformation?
Let us take the example of $ty$~noncommutativity and attempt
to reduce along the $x$~coordinate. 
Simply demanding $\pa_x=0$ in (\ref{res1}) and~(\ref{res2}) yields
\beq
(\unity-P)\,\pa_u\,P \= 0 \und (\unity-P)\,\pa_v\,P \= 0 \quad,
\eeq
which overconstrains the system (unless $\mu{=}0$ or $\mu{=}\infty$).
To remain compatible with the linear system, we must reproduce the $x{=}0$ 
slice of the soliton configuration. Therefore, we put $x=0$ in the coordinate
transformation~(\ref{wdef}), and the reduced coordinates are
\beq\label{zdef}
\bal
z\ &:=\ \nu\,\bigl[\bar{\mu} u +\sfrac{1}{\bar\mu} v\bigr]\ =\
\nu\,\bigl[\sfrac{1}{2}(\bar{\mu}{-}\sfrac{1}{\bar\mu})\,y +
\sfrac{1}{2}(\bar{\mu}{+}\sfrac{1}{\bar\mu})\,t\bigr] \quad,\\[4pt]
\bar z\ &:=\ \bar\nu\,\bigl[\mu u +\sfrac{1}{\mu} v\bigr]\ =\
\bar\nu\,\bigl[\sfrac{1}{2}(\mu{-}\sfrac{1}{\mu})\,y +
\sfrac{1}{2}(\mu{+}\sfrac{1}{\mu})\,t\bigr] \quad,
\eal
\eeq
but this does not mean that $\pa_x$ vanishes! 
Rather, (\ref{parel}) yields the relation
\beq
\pa_x \= \nu\,\pa_z + \bar\nu\,\pa_{\bar z} \quad,
\eeq
which fixes the $x$~dependence in terms of the reduced coordinates.

Time-space noncommutativity for Im$\,\mu^2\neq0$ now takes the form
\beq
[t,y] \= \im\theta
\qquad\Longrightarrow\qquad [z,\bar z] \= 2\,\theta\ >0 \quad,
\eeq
like in the Euclidean Moyal plane.
The reduction along~$x$ is most easily implemented by~(\ref{wvac})
and the relation between $w$ and~$z$:
\beq
w=z+\nu x \quad\ \Longrightarrow\quad\
z|0\> = -\nu x|0\> \quad,
\eeq
and so
\beq
|0\> = \e^{x(\bar\nu z-\nu\bar z)/2\t} |\Omega\> 
\qquad\text{with}\quad z|\Omega\> = 0 \quad.
\eeq
Therefore, expressing the ``$w$~vacuum'' $|0\>$ 
through the ``$z$~vacuum'' $|\Omega\>$
produces a specific phase factor linear in $x$ as well as in the reduced
coordinates. Further $x$~dependence, tied to the moduli~$\a$, 
is seen to cancel in~(\ref{Phisol}). 
Hence, all $x$~dependence in our soliton factorizes, 
and the reduction takes the twisted form
\beq \label{Phired}
\Phi(x,u,v) 
\= \e^{x(\bar\nu z-\nu\bar z)/2\t}\,g(u,v)\,\e^{-x(\bar\nu z-\nu\bar z)/2\t} 
\= \e^{\im x f(u,v)}\,g(u,v)\,\e^{-\im x f(u,v)} 
\eeq
with $f(u,v):=(\bar\nu z{-}\nu\bar z)/2\im\t$. 
One quickly verifies that
\beq \label{deriv}
\pa_x\Phi \= \sfrac{1}{2\theta}\,[\bar\nu z-\nu\bar z\,, \Phi] \=
(\nu\pa_z + \bar\nu\pa_{\bar z})\Phi 
\eeq 
as it should be.
Our soliton configuration~(\ref{Phisol}) cleanly descends to
the two-dimensional unitary configuration
\beq \label{gsol}
g(u,v) \= \e^{\a\bar z-\bar\a z}
\Bigl[ \unity_{\cH} - (1{-}\sfrac{\mu}{\bar\mu})|\Omega\>\<\Omega| \Bigr]
\e^{-\a\bar z+\bar\a z} \quad.
\eeq

{}From its star-product form, obtained by simply putting $x{=}0$ in
(\ref{starsol}), we obtain again a Gaussian profile, but now peaked at
\beq
\bal
y &\= -2\t\,\bigl( \sfrac{\mu}{\bar\mu}{-}\sfrac{\bar\mu}{\mu} \bigr)^{-1}
\bigl( \sfrac{\mu\a}{\nu}{-}\sfrac{\bar\mu\bar\a}{\bar\nu}
{+}\sfrac{\a}{\mu\nu}{-}\sfrac{\bar\a}{\bar\mu\bar\nu} \bigr)\quad,\\
t &\= +2\t\,\bigl( \sfrac{\mu}{\bar\mu}{-}\sfrac{\bar\mu}{\mu} \bigr)^{-1}
\bigl( \sfrac{\mu\a}{\nu}{-}\sfrac{\bar\mu\bar\a}{\bar\nu}
{-}\sfrac{\a}{\mu\nu}{+}\sfrac{\bar\a}{\bar\mu\bar\nu} \bigr)\quad.
\eal
\eeq
Thus, what we have constructed is not a soliton, but rather a 
noncommutative {\it instanton\/}. It is indeed the $x{=}0$ slice 
of the $2{+}1$~dimensional soliton constructed above,
and thus has lost one of its world-volume dimensions.
The commutative limit is singular as had to be expected.

Other choices of the deformed spatial direction confirm the generic picture:
Putting to zero the commutative coordinate yields a time-space noncommutative
integrable model in $1{+}1$ dimensions and slices our one-soliton solution
orthogonal to it, producing an instanton configuration. To obtain 
from~(\ref{Phisol}) a noncommutative {\it soliton\/} in $1{+}1$ dimensions 
one would have to slice it parallel to the velocity,
i.e.~to take the velocity in the deformed direction.\footnote{
Another possibility is to reduce an extended-wave solution of the Ward model
\cite{bieling} along its spatial extension. However, these occur only in
the nonabelian model and descend to the sine-Gordon solitons~\cite{ncsg}.}
However, this is precisely
the singular situation encountered earlier, where only the star-product
formulation is available. No soliton solutions are known for this case.

\section{A new integrable equation}
\noindent
Which is the field equation that governs the reduced configurations?
Inserting our reduction ansatz~(\ref{Phired}) 
into the Ward equation~(\ref{wardeq3}) yields
\beq \label{wardred}
\pa_v\bigl(g^\+\pa_u g\bigr)\ +\ \bigl[ f\,,\,g^\+[f,g]\bigr] \= 0 \quad.
\eeq
Rewriting the commutator term with the help of
\beq
[ f\,,\,.\,] \= \sfrac{1}{2\im\t}\,[ \bar\nu z - \nu\bar z\,,\,.\,]
\= \sfrac{1}{\im}\,(\nu\pa_z + \bar\nu\pa_{\bar z})
\eeq
and also converting the first term via
\beq
\pa_u \= \nu\bar\mu \pa_z + \bar\nu\mu \pa_{\bar z} \und
\pa_v \= \frac{\nu}{\bar\mu} \pa_z + \frac{\bar\nu}{\mu} \pa_{\bar z} \quad,
\eeq
we arrive (after dividing by $\nu\bar\nu$) at
\beq
(\sfrac{\mu}{\bar\mu}{-}1)\,\pa_z\bigl(g^\+\pa_{\bar z}g\bigr)\ +\ 
(\sfrac{\bar\mu}{\mu}{-}1)\,\pa_{\bar z}\bigl(g^\+\pa_zg\bigr) \= 0 \quad.
\eeq
Another way to obtain this equation is to express the Ward 
equation~(\ref{wardeq3}) in the co-moving coordinates,
\beq
(\sfrac{\mu}{\bar\mu}{-}1)\,\pa_w\bigl(\Phi^\+\pa_{\bar w}\Phi\bigr)\ +\
(\sfrac{\bar\mu}{\mu}{-}1)\,\pa_{\bar w}\bigl(\Phi^\+\pa_w\Phi\bigr) \= 0\quad,
\eeq
and to simply pass to the reduced coordinates~$(z,\bar z)$.

The new equation can be simplified to
\beq \label{wardnew}
\mu\,\pa_z\bigl(g^\+\pa_{\bar z}g\bigr)\ -\
\bar\mu\,\pa_{\bar z}\bigl(g^\+\pa_z g\bigr) \= 0 
\eeq
or, parameterizing $\ g=\e^{\im\phi}\ $ by 
a noncommutative real scalar field~$\phi(u,v)$, be written as
\beq
\mu\,\pa_z\bigl( \e^{-\im\phi}\,\pa_{\bar z}\e^{\im\phi} \bigr)\ -\
\bar\mu\,\pa_{\bar z}\bigl( \e^{-\im\phi}\,\pa_z \e^{\im\phi} \bigr)\= 0\quad.
\eeq

It is instructive to interpret this as a sigma model with metric~$h$ 
(symmetric) and magnetic field~$b$ (antisymmetric) by comparing with
\beq
h^{ij}\,\pa_{(i}\bigl(g^\+\pa_{j)} g\bigr)\ +\ 
b^{ij}\,\pa_{[i}\bigl(g^\+\pa_{j]} g\bigr) \= 
(h^{ij}+b^{ij})\,\pa_i\bigl(g^\+\pa_j g\bigr) \= 0
\qquad\text{for}\quad i,j=z,\bar z \quad.
\eeq
Explicitly adding the contributions from the first and second term 
in~(\ref{wardred}), we have
\beq
\bal
\begin{pmatrix} 
h^{zz} & h^{z\bar z} \\[6pt] h^{\bar z z} & h^{\bar z\bar z}
\end{pmatrix} &\= 
\begin{pmatrix}
\nu^2&\frac{\nu\bar\nu}{2}(\frac{\mu}{\bar\mu}{+}\frac{\bar\mu}{\mu})\\[6pt]
\frac{\nu\bar\nu}{2}(\frac{\mu}{\bar\mu}{+}\frac{\bar\mu}{\mu}) & \bar\nu^2
\end{pmatrix} \ -\ 
\begin{pmatrix}
\nu^2 & \ \nu\bar\nu \\[6pt] \nu\bar\nu & \ \bar\nu^2 
\end{pmatrix} \=
\nu\bar\nu\,\frac{(\mu{-}\bar\mu)^2}{2\mu\bar\mu}
\begin{pmatrix} \ 0 & \quad 1 \ \\[6pt] \ 1 & \quad 0 \ \end{pmatrix} \quad,
\\[8pt]
\begin{pmatrix}
b^{zz} & b^{z\bar z} \\[6pt] b^{\bar z z} & b^{\bar z\bar z}
\end{pmatrix} &\=
\frac{\nu\bar\nu}{2}\,\Bigl(\frac{\mu}{\bar\mu}{-}\frac{\bar\mu}{\mu}\Bigr)
\begin{pmatrix} \!\phantom{-}0 & \quad 1\ \\[6pt] \!-1 & \quad 0\ \end{pmatrix}
\ +\ \begin{pmatrix} \,0 & \quad 0 \,\\[6pt] \,0 & \quad 0 \,\end{pmatrix} \=
\nu\bar\nu\;\frac{\mu^2{-}\bar\mu^2}{2\mu\bar\mu}
\begin{pmatrix} \!\phantom{-}0 & \quad 1\ \\[6pt] \!-1 & \quad 0\ \end{pmatrix}
\quad,
\eal
\eeq
showing that the commutator term is of rank one only but serves to cancel
the diagonal metric contribution.
At the same time, it modifies the original Minkowski metric to a Euclidean one,
as is most transparent in the $(t,y)$ basis:
\beq
\begin{pmatrix} h^{tt} & h^{ty} \\[6pt] h^{yt} & h^{yy} \end{pmatrix} \=
\frac{-\mu\bar\mu}{(\mu{+}\bar\mu)^2}
\begin{pmatrix}
|\frac{1}{\mu}{-}\mu|^2 & \ |\frac{1}{\mu}|^2 - |\mu|^2 \\[6pt]
|\frac{1}{\mu}|^2 - |\mu|^2 & \ |\frac{1}{\mu}{+}\mu|^2 
\end{pmatrix}
\und
\begin{pmatrix} b^{tt} & b^{ty} \\[6pt] b^{yt} & b^{yy} \end{pmatrix} \=
\begin{pmatrix} \!\phantom{-}0 & \quad 1\ \\[6pt] \!-1 & \quad 0\ \end{pmatrix}
\quad,
\eeq
while the first term in~(\ref{wardred}) provides only $h=\sigma_3$ 
and $b=\im\sigma_2$. 
In these coordinates, our integrable equation~(\ref{wardnew}) then reads
\beq
\bal
& |\sfrac{1}{\mu}{-}\mu|^2\,\pa_t (g^\+\pa_t g)\ +\
\bigl(|\sfrac{1}{\mu}|^2 - |\mu|^2 - |1{+}\sfrac{\mu}{\bar\mu}|^2\bigr)\,
\pa_t (g^\+\pa_y g)\ +\ {}\\[4pt] {}\ +\ 
& \bigl(|\sfrac{1}{\mu}|^2 - |\mu|^2 + |1{+}\sfrac{\mu}{\bar\mu}|^2\bigr)\,
\pa_y (g^\+\pa_t g)\ +\
|\sfrac{1}{\mu}{+}\mu|^2\,\pa_y (g^\+\pa_y g) \= 0 \quad.
\eal
\eeq
It inherits a hidden global U(1) invariance from the global phase symmetry
$z\mapsto\e^{\im\gamma}z$ of~(\ref{wardnew}).
To summarize this paragraph, the integrable reduction~(\ref{Phired})
effectively not only transforms the metric but also flips its signature!
It is easily checked that this phenomenon does not depend on the spatial
direction chosen for the integrable reduction.

Our novel equation~(\ref{wardnew}) may also be obtained from an action 
principle. For the reduction~(\ref{Phired}), the Nair-Schiff sigma-model-type 
action~\cite{NaS,losev} for SDYM in $2{+}2$ dimensions descends to
\bea
S[g] &\=& -\int\!\diff t\,\diff y\ h^{z\bar z} \,
(g^\+ \pa_z g)\,(g^\+\pa_{\bar z}g) 
\ -\ \int\!\diff t\,\diff y\,\diff\lambda\ b^{z\bar z}\,
\bigl[ \tilde g^\+ \pa_z \tilde g\,,\,\tilde g^\+ \pa_{\bar z}\tilde g \bigr]\,
\tilde g^\+ \pa_\lambda \tilde g \\
&\=& \phantom{-}\int\!\diff t\,\diff y\ \bigl( h^{z\bar z}\,
\pa_z\phi\,\pa_{\bar z}\phi
+ \sfrac{\im}{3} b^{z\bar z}\,[\pa_z\phi\,,\,\pa_{\bar z}\phi]\,\phi 
+ O(\phi^4) \bigr) \quad,
\eea
with a  unitary extension $\tilde g(u,v,\lambda)$ interpolating between
\beq
\tilde g(u,v,0) \= \unity_{\cH} \und
\tilde g(u,v,1) \= g(u,v) \quad.
\eeq
Since the metric~$h$ is Euclidean, the elementary excitation~$\phi$
does not propagate: its mass-shell condition for $(E,p)\equiv(p_t,p_y)$ is
\beq
E \= c\,p \qquad\text{with}\qquad
c \= \frac{\mu^2+1}{\mu^2-1}\quad\text{or}\quad\frac{\bar\mu^2+1}{\bar\mu^2-1}
\eeq
being complex. Nevertheless, the model features instanton-like classical
configurations like~(\ref{gsol}) as we have seen.

It is rather obvious that the solution strategy of section~3
(linear system and single-pole ansatz) applies to~(\ref{wardnew})
with the logical adjustments, i.e.
\beq
w \ \mapsto\ z \und
\pa_x \ \mapsto\ \im [f\,,\,.\,] \= \nu\pa_z + \bar\nu\pa_{\bar z} \quad.
\eeq
On this road one again arrives at the abelian noncommutative instanton
(\ref{gsol}) in the rank-one case.
Higher-rank projectors in~(\ref{TandP}) yield multi-soliton and multi-instanton
solutions in $2{+}1$ and $2{+}0$ dimensions, respectively, with individual
position moduli~$\a^i$ but common velocity given by~$\mu$.
More general, multi-pole, ans\"atze for~$\psi$ should produce multi-solitons
with {\it relative\/} motion on time-space noncommutative~$\R^{2,1}$, but their
integrable reduction to two dimensions is unclear because the reduction
(\ref{Phired}) depends on~$\mu$. These configurations will be genuinely
$2{+}1$ dimensional with time-space noncommutativity.

Finally we remark that our twisted reduction ansatz (\ref{Phired}) 
extends straightforwardly for the nonabelian case. With an~$f$ 
of diagonal form, one can embed into the nonabelian group many abelian 
solitons whose generic features have been examined and discussed above.

\bigskip

\noindent
{\bf Acknowledgements\ }\\
The authors would like to thank the organizers of the Bayrischzell Workshop  
2005 for organizing a stimulating meeting, where the main ideas of this work 
were conceived. CSC thanks T.~Inami for helpful discussions on integrable 
systems. OL thanks A.D.~Popov for helpful comments.
CSC acknowledges the support of EPSRC through an advanced fellowship.
The work of OL is partially supported by the Deutsche Forschungsgemeinschaft
(DFG).

\end{document}